\documentclass[twocolumn,showpacs,preprintnumbers,amsmath,amssymb]{revtex4}

\usepackage{graphicx}
\usepackage{color}
\usepackage{dcolumn}
\usepackage{bm}

\begin{document}
\preprint{}

\title{The role of long waves in the stability of the plane wake}

\author{Stefania Scarsoglio}
\affiliation{%
Dipartimento di Ingegneria
Aeronautica e Spaziale, Politecnico di Torino,10129 Torino, Italy\\
International Center for Turbulence Cooperation, ICTR.}%
\author{Daniela Tordella}%
 \email{daniela.tordella@polito.it}
\affiliation{%
Dipartimento di Ingegneria
Aeronautica e Spaziale, Politecnico di Torino,10129 Torino, Italy\\
International Center for Turbulence Cooperation, ICTR.}%
\author{William O. Criminale}
\affiliation{
Department of Applied Mathematics,
University of Washington, Seattle, WA 98195-2420, USA}%

\date{\today}
%
\begin{abstract}
This work is directed towards investigating the fate of three-dimensional long perturbation waves in a plane incompressible wake.
The analysis is posed as an initial-value problem in space. More specifically, input is made at an initial location
in the downstream direction and then tracing the resulting behavior further downstream subject to the restriction of finite kinetic energy. 
This presentation 
follows the outline given by Criminale and Drazin [Stud.
in Applied Math. \textbf{83}, 123 (1990)] that describes the system
in terms of perturbation vorticity and velocity.   The analysis is based on large scale waves
and expansions using multi scales and multi times for the partial differential equations. The multiscaling is based on an approach where the small parameter is
linked to the perturbation property independently from the flow control parameter. Solutions of the perturbative equations are determined  numerically after the introduction of a regular perturbation scheme analytically deduced up to the second order. Numerically, the complete linear system is also integrated. Since the results relevant to the complete problem are in very good agreement with the results of the first order analysis, the numerical solution at the second order was deemed not necessary. The use for an arbitrary initial-value problem will be
shown to contain a wealth of information for the  different transient behaviors associated to the symmetry, angle of obliquity and spatial decay of the long  waves. The amplification factor of transversal perturbations never presents the trend -- a growth followed by a long damping - usually seen in waves with wavenumber of order one or less. Asymptotical instability is always observed.
\end{abstract}

\maketitle

\section{Introduction}

The traditional investigation of stability of shear flows is cast as a linear initial-value perturbation
problem. In principle, save for the additional complexity of necessitating three space dimensions as
well as time, this is done by means of a Laplace transform in time. Once the boundary conditions have
been satisfied, the stability or non stability is found. Further, depending upon the mean shear flow that is
being investigated, the causes are determined. No attention is given to any specific input or the effect  of
various physics.  Moreover, little attention was given to early period
dynamics, see for example Grosch and Salwen \cite{GS68}, Salwen and Grosch \cite{SG81}. These authors showed that
there can be early time growth of a perturbation even if there is damping for long time. In short, a branch
cut can be present as well as any pole when inverting the Laplace transform.
Next, from the laboratory, interest turned to spatial growth or decay after an input at an initial location
rather than the temporal behavior. This construction creates new difficulties but they are not insurmountable
\cite{CJJ2003}, \cite{SH2001}. Still, just as in the temporal problem, no specific initial input has been examined.
Regardless of the framework, it has been known since the first results of stability theory, that the value
of the wavenumber that comes into the analytical framework (due to Fourier decomposition in the variables
in the plane that is perpendicular to the mean flow) is small in the regions where there is
instability.  In short, long waves.  Such a result provides a sound means for the analysis and examination
of a specific initial input. This is true whether posed as a temporal or spatial initial-value problem. It
further provides a means to investigate interaction, the early period and a way to suppress any growth
at the early period or location.
In 1962 a study about the instability to long waves of unbounded parallel inviscid flow was given by Drazin and Howard \cite{DH62}. Using the normal mode analysis, they found that there is a finite number of different modes unstable to long waves, essentially one for each relative maximum or minimum of the velocity profile. Healey \cite{H06} considered long waves for investigating spatial instability of the rotating-disc boundary layer, and by means of an analytic theory in the inviscid long wave limit, he obtained an explicit expression for the growth rate in terms of  basic flow parameters.

Large or long waves have now been used in full
nonlinear simulations. For example, Ryzkov and Shevtsova \cite{RS09} focused on convective instability in  multicomponent fluids, showing by means of both linear stability analysis and nonlinear numerical calculations that the instability is caused by the interplay between the basic flow and the concentration waves which have a long scale in a vertical direction. And Barros and Choi \cite{BC09} considered the inhibition of the shear instability that can be induced by large amplitude internal solitary waves travelling in a two-layer flow with a top free surface.
For large eddy simulations in turbulence see \cite{PZSH90}
or \cite{GPMC91} among many others.

The analysis  in the present work is based on large scale waves
and expansions using multi scales and multi times for the partial differential equations. The multiscale is based on an approach where the small parameter is linked to the perturbation property  independently from the flow control parameter. In fact, the perturbation scheme is based on the introduction of a small parameter which is the wavenumber $k$ in the limit $k \rightarrow 0$, and is analytically determined to the second order. The perturbative equations used follow the formulation given by Criminale and Drazin \cite{CD90} that describes the system in terms of perturbation vorticity and velocity.
Numerically, the complete linear system has also been integrated for the non parallel base flow. What results is an extension of a previous work based on a locally near parallel assumption \cite{STC09}.
The formulation of the linear perturbation initial-value problem is presented in Section 2.
Results are in Section 3. Conclusions follow in Section 4.

\section{Formulation}

By exciting the plane wake flow ({\bf{U}}=$(U(x,y; Re), V(x,y; Re))$) with small arbitrary
three-dimensional perturbations, the continuity and Navier-Stokes
equations for the perturbed system  linearized with respect to small
oscillations are given by

\begin{equation}
\frac{\partial \widetilde{u}}{\partial x} + \frac{\partial
\widetilde{v}}{\partial y} + \frac{\partial
\widetilde{w}}{\partial z} = 0 \label{IVPMultiscale_continuity}
\end{equation}
\begin{equation}
 \frac{\partial \widetilde{u}}{\partial t} + \widetilde{u} \frac{\partial U}{\partial x} + U \frac{\partial
\widetilde{u}}{\partial x} + \widetilde{v} \frac{\partial
U}{\partial y} +  V \frac{\partial \widetilde{u}}{\partial y} +
\frac{\partial \widetilde{p}}{\partial x} = \frac{1}{Re}
\nabla^2\widetilde{u} \label{IVPMultiscale_NS1}
\end{equation}
\begin{equation}
 \frac{\partial \widetilde{v}}{\partial t} + \widetilde{u} \frac{\partial V}{\partial x} + U \frac{\partial
\widetilde{v}}{\partial x} +  \widetilde{v} \frac{\partial
V}{\partial y} +  V \frac{\partial \widetilde{v}}{\partial y} +
\frac{\partial \widetilde{p}}{\partial y} = \frac{1}{Re}
\nabla^2\widetilde{v} \label{IVPMultiscale_NS2}
\end{equation}
\begin{equation}
 \frac{\partial \widetilde{w}}{\partial t} + U \frac{\partial
\widetilde{w}}{\partial x} +  V \frac{\partial
\widetilde{w}}{\partial y} + \frac{\partial
\widetilde{p}}{\partial z} = \frac{1}{Re} \nabla^2\widetilde{w}
\label{IVPMultiscale_NS3}
\end{equation}

\noindent where ($\widetilde{u}(x, y, z, t)$, $\widetilde{v}(x, y,
z, t)$, $\widetilde{w}(x, y, z, t)$) and $\widetilde{p}(x, y, z, t)$
are the components of the perturbation velocity and pressure,
respectively.

\begin{figure}[t]
  \center
  \includegraphics[width=\columnwidth]{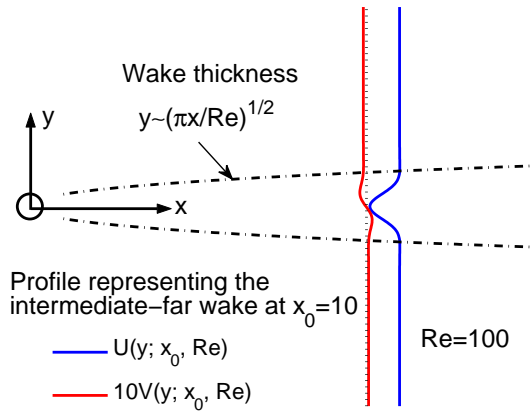}\\
  \caption{Base flow sketch. The base flow has been chosen in order to be an acceptable  representation of the intermediate-far field. To this aim we build a homogeneous field  in $x,z$  by using the information associated to a section, $x_0$,  placed in the intermediate region, $x \in [5,30]$. In the sketch the longitudinal and transversal
  profiles at $Re=100$ are frozen at $x_0=10$ (note that the
  transversal velocity V is multiplied by a factor $10$). The base flow ($U(y;x_0,Re), V(y;x_0,Re)$)
  is thus a slightly non parallel
  flow homogeneous in $x,z$, which makes it possible to Laplace transform the
  perturbative equations
  in  $x$ and to Fourier transform them in $z$.}\label{base_flow_scheme}
\end{figure}

\begin{figure}
\begin{minipage}[]{0.6\columnwidth}
   \includegraphics[width=\columnwidth]{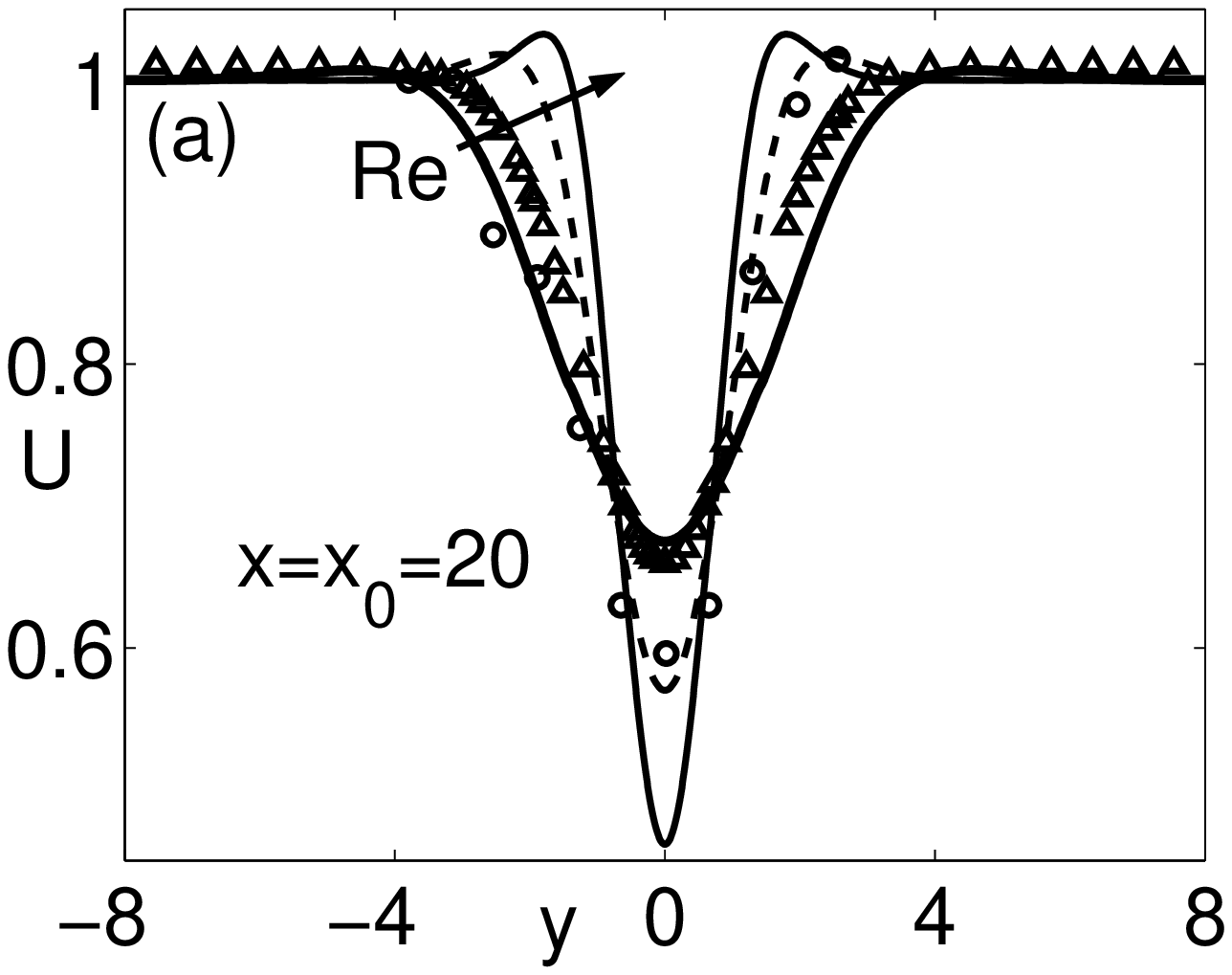}
    \label{U_x20}
\end{minipage}
\begin{minipage}[]{0.6\columnwidth}
   \includegraphics[width=\columnwidth]{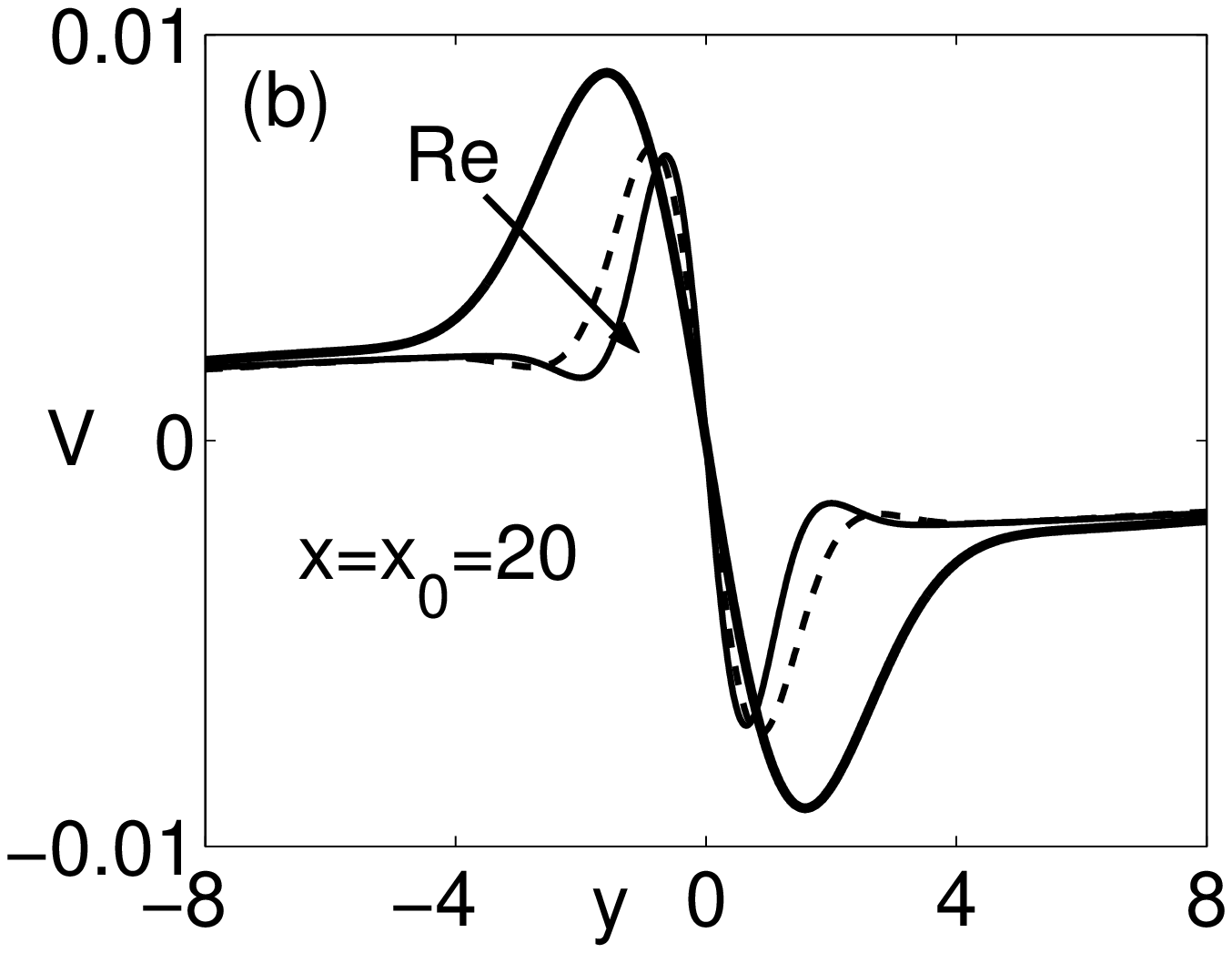}
    \label{V_x20}
\end{minipage}
\caption{Example of velocity profiles in the intermediate wake at the
downstream station $x = x_0 = 20$. (a) Longitudinal velocity $U$, (b) transversal velocity
$V$. Continuous curves: analytical solutions ($Re = 20, 60$ and $100$) by Tordella and Belan (2003) \cite{TB03}, triangles: numerical results ($Re = 34$) by Berrone (2001) \cite{B01}, circles: laboratory data ($Re = 34$)
by Kovasznay (1948) \cite{K48}.}\label{UW_x20}
\end{figure}

The independent spatial variables $z$ and $y$ are defined from
$-\infty$ to $+\infty$, $x$ from $0$ to $+\infty$. All physical
quantities are normalized with respect to the free stream velocity $U_f$,
the body scale $D$ and the density $\rho$. 
The Reynolds number is defined as $Re=U_f D/\nu$, where $\nu$ is the
kinematic viscosity.

The two dimensional wake is a thin free flow that can be schematized as shown in Fig.\ref{base_flow_scheme}. Leaving aside the near field, that is highly non parallel since it hosts the two symmetric counter circulating vortices that constitute the separation region, the intermediate and long term wake is a near parallel flow. The wake slowly becomes thicker according to a law which, at first order, scales as $\displaystyle{\left (\frac{x}{Re}\right)^{\frac 1 2}}$.  As  representation  of this steady subcritical flow we consider  the asymptotic expansion solution in inverse powers of  $x$ obtained in \cite{TB03}. In particular,  we consider the intermediate far field  well represented by a section placed near $x= x_0 = 10$ and  build the basic flow by freezing it at this longitudinal station. In so doing, the basic  flow is  parameterized through the
downstream station $x_0$ and the Reynolds number $Re$ ({\bf{U}}=$(U(y; x_0, Re), V(y; x_0, Re))$). It is thus homogeneous in $x$ and $z$. As such, the long waves that are the main subject of this study are valid.

The explicit expressions of the base
flow longitudinal and transversal components are:

\begin{eqnarray}
\nonumber U(y; x_0, Re) & = & \phi_0 + \phi_1 x_0^{-1/2} \\ & & + \phi_2 x_0^{-1} + \phi_3 x_0^{-3/2}   \label{u3} \\
\nonumber V(y; x_0, Re) & = & \chi_0 + \chi_1 x_0^{-1/2} \\ & & +
\chi_2 x_0^{-1} + \chi_3 x_0^{-3/2}  \label{v3}
\end{eqnarray}

The coefficients $\phi_i$ = $\phi_i(y; x_0, Re)$  and $\chi_i$ = $\chi_i(y; x_0, Re)$  of this expansion up to $i=3$  are given in Appendix A. Fig. 2 displays the intermediate wake profile for which there exists a comparison based on laboratory and numerical simulation results, see also
\cite{TB03},\cite{B01},\cite{K48}, \cite{TS09}.

It should be noted that when using such a kind of representation the base flow
nonparallelism is considered and allows for the effect of the
lateral entrainment to be obtained \cite{TS09}. However, in this
study a fixed location $x_0$ of the intermediate wake is considered
since in this region absolute instability pockets have been found by
recent modal analyses \cite{TSB06}, \cite{BT06}. The term intermediate is used in the
general sense as that given by Barenblatt \cite{B66}: '...
intermediate asymptotics are self-similar or near-similar solutions
of general problems valid for times and distances from boundaries,
large enough for the influence of the fine details of the initial or
boundary conditions to be insignificant, but small enough that the
system is far from the ultimate equilibrium state...'. The distance
beyond which the intermediate region is assumed to begin varies from
eight to four diameters $D$ for $Re \in [20, 40]$ \cite{TB03}.

By combining the momentum equations
(\ref{IVPMultiscale_NS1}) to (\ref{IVPMultiscale_NS3}) to eliminate
the pressure, the resulting governing equations become

\begin{eqnarray}
\frac{\partial \nabla^2 \widetilde{v}}{\partial t} &=&  - [U
\frac{\partial}{\partial x} + V \frac{\partial}{\partial y} -
\frac{1}{Re} \nabla^2] \nabla^2 \widetilde{v} - [a \frac{\partial}{\partial x} + \nonumber \\
&& + b] \widetilde{u} -
[\frac{\partial \Omega_z}{\partial y} \frac{\partial}{\partial x}
+ c(y)] \widetilde{v} + a \frac{\partial
\widetilde{w}}{\partial z} + \nonumber \\
&& + \frac{\partial V}{\partial y}
\frac{\partial \widetilde{\omega}_x}{\partial z} - e \frac{\partial \widetilde{\omega}_y}{\partial z} - [d \frac{\partial}{\partial x} +
e \frac{\partial}{\partial y}] \widetilde{\omega}_z,\label{IVPmultiscale_OS1} \\
\frac{\partial \widetilde{\omega}_y}{\partial t} &=& - [U
\frac{\partial}{\partial x} + V \frac{\partial}{\partial y} +
d - \frac{1}{Re} \nabla^2]
\widetilde{\omega}_y + \nonumber \\
&& - \frac{\partial U}{\partial y} \frac{\partial
\widetilde{v}}{\partial z} + e \frac{\partial \widetilde{w}}{\partial y},
\label{IVPmultiscale_SQUIRE1}
\end{eqnarray}

\noindent where $(\widetilde{\omega}_x,
\widetilde{\omega}_y, \widetilde{\omega}_z)$ is the perturbation
vorticity field, $\displaystyle{\Omega_z = ( \frac{\partial V}{\partial x}  -  \frac{\partial
U}{\partial y})  |_{x=x_0}}$ is the mean vorticity in the spanwise direction, and the coefficients $a$, $b$, $c$, $d$, $e$ are the spatial derivatives of the base flow vorticity and velocity at $x_0$, namely: 

\noindent $\displaystyle{a= \left. \frac{\partial \Omega_z}{\partial x} \right |_{x=x_0}}$, $\displaystyle{b= \left.
\frac{\partial^2 \Omega_z}{\partial x^2} \right |_{x=x_0}}$,
$\displaystyle{c(y)= \left. \frac{\partial^2 \Omega_z}{\partial x \partial y} \right |_{x=x_0}}$, $\displaystyle{d= \left. \frac{\partial U}{\partial x} \right |_{x=x_0}}$, $\displaystyle{e= \left. \frac{\partial V}{\partial x} \right |_{x=x_0}}$.

By introducing the quantity
$\widetilde{\Gamma}$, that is defined by
\begin{equation}
\nabla^2\widetilde{v} = \widetilde{\Gamma} \label{Vel_Vor}
\end{equation}
\noindent we obtain three coupled equations
(\ref{IVPmultiscale_OS1}), (\ref{IVPmultiscale_SQUIRE1}) and
(\ref{Vel_Vor}). Equations (\ref{IVPmultiscale_OS1}) and
(\ref{IVPmultiscale_SQUIRE1}) are the Orr-Sommerfeld and Squire
equations respectively, from the classical linear stability analysis
for three-dimensional disturbances. From kinematics, the relation
\begin{equation}
\widetilde{\Gamma} = \frac{\partial \widetilde{\omega}_z}{\partial
x} - \frac{\partial \widetilde{\omega}_x}{\partial z}
\label{Vel_Vor_Kin}
\end{equation}
\noindent physically links the
perturbation vorticity components in the $x$ and $z$ directions
($\widetilde{\omega}_x$ and $\widetilde{\omega}_z$ respectively) and
the perturbed velocity field. By combining equations
(\ref{IVPmultiscale_OS1}) and (\ref{Vel_Vor}) then
\begin{eqnarray}
\frac{\partial \widetilde{\Gamma}}{\partial t} &=&  - [U
\frac{\partial}{\partial x} + V \frac{\partial}{\partial y} -
\frac{1}{Re} \nabla^2] \widetilde{\Gamma} - [a \frac{\partial}{\partial x} + \nonumber \\
&& + b] \widetilde{u} -
[\frac{\partial \Omega_z}{\partial y} \frac{\partial}{\partial x}
+ c(y)] \widetilde{v} + a \frac{\partial
\widetilde{w}}{\partial z} + \nonumber \\
&&  + \frac{\partial V}{\partial y}
\frac{\partial \widetilde{\omega}_x}{\partial z} - e \frac{\partial \widetilde{\omega}_y}{\partial z} - [d \frac{\partial}{\partial x} +
e \frac{\partial}{\partial y}]
\widetilde{\omega}_z, \label{OS_Gamma}
\end{eqnarray}
\noindent which, together with
(\ref{IVPmultiscale_SQUIRE1}) and (\ref{Vel_Vor}), fully describes
the perturbed system. \noindent Since seven unknown quantities
($\widetilde{u}, \widetilde{v}, \widetilde{w}, \widetilde{\omega}_x,
\widetilde{\omega}_y, \widetilde{\omega}_z, \widetilde{\Gamma}$) are
involved in the above equations (\ref{IVPmultiscale_SQUIRE1}),
(\ref{Vel_Vor}) and (\ref{OS_Gamma}), the perturbation vorticity
definition and the continuity equation

\begin{eqnarray}
\widetilde{\underline{\omega}} &=& \nabla \times
\widetilde{\underline{u}}, \label{IVPmultiscale_vort} \\
\nabla \cdot \widetilde{\underline{u}} &=& 0,
\label{IVPmultiscale_cont}
\end{eqnarray}

\noindent link the perturbative system of equations
(\ref{IVPmultiscale_SQUIRE1}), (\ref{Vel_Vor}) and
(\ref{OS_Gamma}).

\begin{figure}
  \center
  \includegraphics[width=\columnwidth]{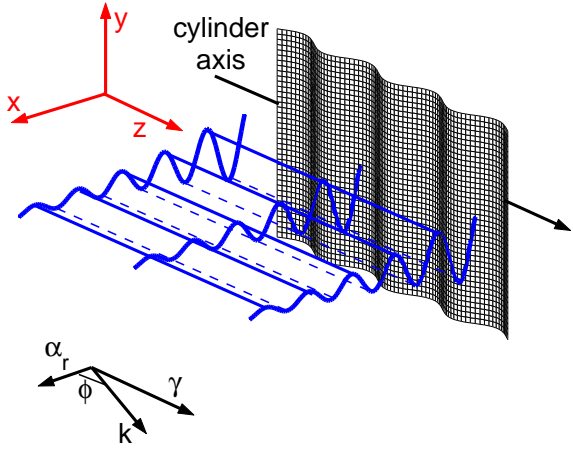}\\
  \caption{Perturbation geometry scheme.}\label{DT_perturbation_scheme}
\end{figure}

\noindent For every dependent variable, we perform a combined
spatial Laplace-Fourier decomposition in the $x$ and $z$ directions
defined by

\begin{equation}\label{IVPMultiscale_Four_Lapl_trans}
\hat{g}(y, t; \alpha, \gamma) = \int_{-\infty} ^{+\infty} \int_{0}
^{+\infty} \widetilde{g}(x, y, z, t) e^{-i \alpha x -i \gamma z} dx
dz \nonumber
\end{equation}

\noindent where  $\widetilde{g}$ is the general dependent variable,
$\alpha$, the longitudinal wavenumber,  is complex ($\alpha =
\alpha_r + i \alpha_i$) and $\gamma$, the transversal wavenumber, is
real. By adopting the velocity-vorticity formulation, \cite{CD90},
\cite{CJJ2003}, the governing equations
(\ref{IVPmultiscale_SQUIRE1}), (\ref{Vel_Vor}) and (\ref{OS_Gamma})
can now be written as

\begin{eqnarray} \label{IVPMultiscale_fou1}
\frac{\partial^2 \hat{v}}{\partial y^2}  &-& (k^2 - \alpha_i^2 + 2
i k cos(\phi) \alpha_i) \hat{v}=
\hat{\Gamma} \\
\label{IVPMultiscale_fou2} \frac{\partial \hat{\Gamma}}{\partial t} &=& \emph{G} \hat{\Gamma} + \emph{H} \hat{v} + \emph{K} \hat{\omega}_y \\
\label{IVPMultiscale_fou3} \frac{\partial \hat{\omega}_y}{\partial
t} &=& \emph{L} \hat{\omega}_y + \emph{M} \hat{v}
\end{eqnarray}

\noindent where $\phi = tan^{-1}(\gamma/\alpha_r)$ is the
perturbation angle of obliquity with respect to the $x$-$y$ physical
plane, $k = \sqrt{\alpha_r^2 + \gamma^2}$ is the polar wavenumber,
$\alpha_i$ is the imaginary part of the complex
longitudinal wavenumber, 
$\hat{\omega}_y$ is the transversal component of the perturbation
vorticity, and $\hat{\Gamma}$ is the vorticity component in the
oblique direction which is defined as $\hat{\Gamma} = i (\alpha
\hat{\omega}_z - \gamma \hat{\omega}_x)$. In Figure
\ref{DT_perturbation_scheme} the three-dimensional perturbative
geometry scheme is shown.

\begin{figure}
  \center
  \includegraphics[width=\columnwidth]{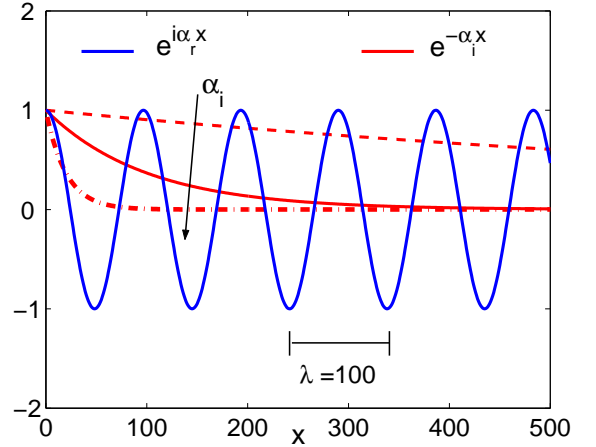}\\
  \caption{The wave spatial evolution in the $x$ direction for k = $\alpha_r = 0.05$, $\alpha_i=0.1, 0.01, 0.001$.}\label{wave_alphai}
\end{figure}

\noindent In order to have a finite perturbation kinetic energy,
$\alpha_i$ can only assume non-negative values. In so doing, we
allow for perturbative waves that can spatially decay ($\alpha_i >
0$) or remain constant in amplitude ($\alpha_i = 0$). In the
following, $\alpha_i$ is called spatial damping rate.
It should be pointed out that the present analysis
is not the standard eigenvalue problem where poles result. Here, in
fact, the spatial damping rate $\alpha_i$ is a parameter and, as such, should be simply imposed. The magnitude of the spatial damping
rate can vary in order to describe a physically meaningful damping
of the perturbative wave in the $x$ direction (disturbances
immediately damped to zero are not allowed). According to this and since long waves
($k\sim10^{-1},10^{-2}$) are considered, $\alpha_i$ is non-negative
and at maximum $\sim 10^{-1}$, see Fig.
\ref{wave_alphai}. \noindent Symbols \emph{G},
\emph{H}, \emph{K}, \emph{L} and \emph{M} represent ordinary
differential operators, written in the form
$\emph{G} = \emph{G}(y; x_0, k, \phi, \alpha_i,
Re)$, and similarly for \emph{H}, \emph{K}, \emph{L} and \emph{M},
since they are functions of $y$ and are parameterized through
the fixed longitudinal station $x_0$, the polar
wavenumber $k$, the angle of obliquity $\phi$, the spatial damping
rate $\alpha_i$ and the Reynolds number $Re$. 
All these operators are explicitly
given in Appendix B.

Equations (\ref{IVPMultiscale_fou1}), (\ref{IVPMultiscale_fou2}) and
(\ref{IVPMultiscale_fou3}) require proper initial and boundary
conditions in order to be solved. As far as the boundary conditions
are concerned, among all solutions, those whose perturbation
velocity field vanishes in the free stream are sought. The initial
conditions are necessary for $\hat{\Gamma}$ and $\hat{\omega}_y$. As
far as the initial conditions for $\hat{\Gamma}$ are concerned,
according to  equation (6),
they can be shaped in terms of set of functions in the $L^2$
Hilbert space via the variable $\hat{v}$, which is here represented by the trigonometric system

$$\hat{v}(0,y) = e^{-y^2} \textmd{cos}(y), \,\,\,\,\,\,\,\,\, \hat{v}(0,y) = e^{-y^2} \textmd{sin}(y),$$

\noindent for the symmetric and the asymmetric perturbations,
respectively. This trigonometrical system is a Schauder basis in
each space $L^p[0,1]$, for $1<p<\infty$. 
The transversal
vorticity $\hat{\omega}_y$ is instead taken initially equal to zero in the
$y$ domain, in order to directly observe the net contribution of
three-dimensionality on its temporal
evolution. However, it can be demonstrated that the eventual
introduction of an initial transversal vorticity does not
substantially affect the perturbation temporal evolution.

In the stability analysis of spatially developing flows, different
scales can be determined. Usually, long and slow scales, related to the
slow base flow evolution, as well as short and fast scales, linked
to the disturbance dynamics, can be defined. However, it should be noticed that in some
flow configurations, long waves can be
destabilizing. Examples of this behavior are the two-dimensional
Blasius boundary layer, the three-dimensional cross-flow
boundary layer, as well as the free shear flows. In such instances, the perturbation wavenumber 
is less than $O(1)$ where instability occurs. Thus, a regular
perturbation scheme can be adopted, defining the polar wavenumber
$k$ as the small parameter \cite{LJJC99}, \cite{S08}.
It should be noted that by using such a
long-wave expansion, the x-scale length of the perturbations is
comparable to the x-scale length of the base flow. Indeed, we only
consider the intermediate and far wake sections, where the flow
slowly evolves in the longitudinal direction. Thus the near wake is
not taken into account. In synthesis,  the  scale of the intermediate wake is of the order
$x_0 \sim 10^1$, and the scale of the long perturbative waves is $\lambda=2 \pi/k
\sim 10^1, 10^2$.

\noindent Two spatial scales, a short one, $y$, and a long one, $Y =
k y$, are defined. For the temporal dynamics, three temporal scales,
the fast one, $t$, and the slow ones, $\tau = k t$ and $T = k^2 t$,
can be identified. The perturbation quantities ($\hat{v},
\hat{\Gamma}, \hat{\omega}_y$) are now function of $y, Y, t, \tau,
T$, expressed as $\hat{\Gamma} = \hat{\Gamma}(y, Y, t, \tau, T; k,
\phi, \alpha_i)$, and similarly for $\hat{v}$ and $\hat{\omega}_y$,
and can be expanded as
\begin{eqnarray}
  \hat{v} &=& \hat{v}_0 + k \hat{v}_1 + k^2 \hat{v}_2 + \cdots, \nonumber \\
  \hat{\Gamma} &=& \hat{\Gamma}_0 + k \hat{\Gamma}_1 + k^2 \hat{\Gamma}_2 + \cdots, \nonumber \\
  \hat{\omega}_y &=& \hat{\omega}_{y0} + k \hat{\omega}_{y1} + k^2 \hat{\omega}_{y2} +
  \cdots \,\,. \label{IVPMultiscale_pert_exp}
\end{eqnarray}
\noindent Initial conditions at order $O(1)$ are defined as in the
full linear problem, while at higher orders ($O(k), O(k^2), \ldots
$) are equal to zero. Boundary conditions remain as stated in the
full linear problem. Substituting relations
(\ref{IVPMultiscale_pert_exp}) in the full linear system
(\ref{IVPMultiscale_fou1}) to (\ref{IVPMultiscale_fou3}), the
following ordered hierarchy of equations, expressed up to $O(k)$,
result and are:
\smallskip

\noindent \textit{\textbf{Order O(1)}}

\begin{eqnarray}
\frac{\partial^2 \hat{v}_0}{\partial y^2} + \alpha_i^2 \hat{v}_0
&=&
\hat{\Gamma}_0 \label{IVPMultiscale_ord1_fou1} \\
\frac{\partial \hat{\Gamma}_0}{\partial t} - G_0 \hat{\Gamma}_0 - H_0 \hat{v}_0 &=& 0 \label{IVPMultiscale_ord1_fou2} \\
\frac{\partial \hat{\omega}_{y0}}{\partial t} - L_0
\hat{\omega}_{y0} &=& 0\label{IVPMultiscale_ord1_fou3}
\end{eqnarray}

\smallskip

\noindent \textit{\textbf{Order O(k)}}

\begin{eqnarray}
\frac{\partial^2 \hat{v}_1}{\partial y^2} &+& \alpha_i^2 \hat{v}_1 = - 2 \frac{\partial^2 \hat{v}_0}{\partial y \partial Y} + 2 i cos(\phi) \alpha_i \hat{v}_0 + \hat{\Gamma}_1  \label{IVPMultiscale_ordk_fou1} \\
\frac{\partial \hat{\Gamma}_1}{\partial t} &-& G_0 \hat{\Gamma}_1 - H_0 \hat{v}_1  = \nonumber\\
  &=&- \frac{\partial \hat{\Gamma}_0}{\partial \tau} + G_{1} \hat{\Gamma}_0 + H_{1} \hat{v}_0 + K_{1} \hat{\omega}_{y0} \label{IVPMultiscale_ordk_fou2}\\
\frac{\partial \hat{\omega}_{y1}}{\partial t} &-& L_0
\hat{\omega}_{y1} = \nonumber\\
 &=&- \frac{\partial \hat{\omega}_{y0}}{\partial
\tau} + L_{1} \hat{\omega}_{y0} + M_{1} \hat{v}_{0}
\label{IVPMultiscale_ordk_fou3}
\end{eqnarray}

\noindent Operators $G_0 = G_0(y; x_0, \phi,
\alpha_i, Re)$ as well as $H_0$ and $L_0$ are functions of the
short scale $y$  only. Operators $G_{1} = G_{1}(y,
Y; x_0, \phi, \alpha_i, Re)$ as well as $H_{1}$, $K_{1}$, $L_{1}$
and $M_{1}$ are function of both the short scale $y$ as well as the
long scale $Y$. These operators are explicitly given in Appendix B.

A comment concerning the role of $\alpha_i$ is needed.
Equations (10) to (15) above are obtained for the case where $\alpha_i >
0$. It can be observed, see the Appendix B, that if $\alpha_i = 0$, the $O(1)$ operators $H_0, L_0$ (but also the $O(k)$ operators $H_1, L_1, M_1, N_1$) are singular. It is possible to verify that if   $\alpha_i = 0$ the disturbances initially imposed remain constant as time passes and reach, in the end, an
asymptotic condition of marginal stability. This fact is
deduced by considering equation (\ref{IVPMultiscale_ord1_fou1}). For
$\alpha_i = 0$, the homogeneous solution assumes the expression
$\hat{v}_{0h} = c_1 + c_2 y$. Since the perturbation velocity field has to vanish in the free stream, $c_1 = 0$ and $c_2 = 0$. Thus
$\hat{v}_0$, and therefore $\hat{\Gamma}_0$, vanish as well.
This means that, in equation (\ref{IVPMultiscale_ord1_fou2}), $\frac{\partial \hat{\Gamma}_0}{\partial t}=0$,
so that there is no temporal evolution for $\hat{\Gamma}_0$. Since
the transversal vorticity $\hat{\omega}_{y0}$ is initially taken as zero, then in equation (\ref{IVPMultiscale_ord1_fou3}),
$\frac{\partial \hat{\omega}_{y0}}{\partial t}=0$, and thus, also for
the transversal vorticity, there is no temporal evolution.
 The complete problem is defined for $\alpha_i = 0$ and, for this value of $\alpha_i$, it does not necessarily show a condition of marginal stability (e.g. see Fig. 6 in the following). However, it is possible to see that the multiscaling limit for $\alpha_i \rightarrow 0$ well approximates the complete problem: cf. Figures 5 and 6. When $\alpha_i = 0$,  the multiscaling presents a discontinuity, since it has a right limit different from the value shown at $\alpha_i = 0$.

Order $O(1)$ is the more important approximation of the
perturbative analysis and its formal expression is simplified with
respect to the complete problem. Numerically, the complete linear system was also integrated. Since the results relevant to the complete problem are in very good agreement with the results of the first order analysis, in the present work, attention is
focused on the resolution of the multiscaling at order $O(1)$. It should be noted that at this order only the short spatial scale $y$ is of relevance.

In the following, a summary of the most significant transient behavior and asymptotic
fate of three-dimensional perturbations is presented to highlight
the agreement between solutions of multiscaling at order $O(1)$
and full linear problem. Results will be principally focused on
parameters such as the spatial damping rate, the polar wavenumber value (to check the validity of the approximation), the angle of obliquity and the
symmetry of the three-dimensional disturbance.

To measure the transient growth the concepts of kinetic energy
density $e(t; k, \phi, \alpha_i)$

\begin{eqnarray}
\nonumber&&e(t; k, \phi, \alpha_i) = \frac{1}{2} \frac{1}{2y_d}
\int_{-y_d}^{+y_d} (|\hat{u}|^2 + |\hat{v}|^2 + |\hat{w}|^2) dy =\\
&&\frac{1}{2} \frac{1}{2y_d} \frac{1}{|k^2 + 2 i k cos(\phi)
\alpha_i - \alpha_i^2|} \times \label{IVPenergydensity}\\ && \times \int_{-y_d}^{+y_d}
(|\frac{\partial \hat{v}}{\partial y}|^2 + |k^2 + 2 i k cos(\phi)
\alpha_i - \alpha_i^2| |\hat{v}|^2 + |\hat{\omega}_y |^2) dy
\nonumber,
\end{eqnarray}

\noindent and normalized amplification factor $G(t; k, \phi,
\alpha_i)$

\begin{equation}
G(t; k, \phi, \alpha_i) = \frac{e(t; k, \phi, \alpha_i)}{e(t=0; k,
\phi, \alpha_i)}.
\end{equation}

\noindent are introduced for both multiscale $O(1)$ quantities
($\hat{v}_0, \hat{\Gamma}_0, \hat{\omega}_{y0}$) and full
problem quantities ($\hat{v}, \hat{\Gamma}, \hat{\omega}_{y}$).

In (16), the limits
$\pm y_d$ define the spatial extension of the  numerical domain. The value
$y_d$ is defined so that the numerical solutions are insensitive
to further extensions of the computational domain size. Here, in
the limit of long waves,  the size of the spatial
domain $2 y_d$  assumes values in the range between $30$ and
$100$ external flow scale $D$. The total kinetic energy can be obtained by integrating the
energy density over all $k$ and $\phi$.

To evaluate the asymptotic behavior we introduce the
temporal growth rate $r$, defined as

\begin{equation}
r(t; k, \phi, \alpha_i) = \frac{log|e(t; \alpha, \gamma)|}{2t},
\;\;\; t>0. \label{IVP2_tgr}
\end{equation}

\noindent The temporal growth rate $r$ is not defined for $t = 0$.
This quantity has, in fact, a precise physical meaning
asymptotically in time.  Moreover,  for both multiscale  and the full problem solutions, the angular frequency
(pulsation) $\omega$ of the perturbation can be introduced by
defining a local, in space and time, time phase $\varphi$ of the
complex wave

\begin{eqnarray}
\hat{v}(y, t; \alpha, \gamma, Re) = A_t(y; \alpha, \gamma, Re) e^{i \varphi(t)}
\end{eqnarray}

\noindent and then computing the time derivative of the phase
perturbation $\varphi$

\begin{eqnarray}
\omega(y, t; k, \phi, \alpha_i) = \frac{d \varphi(y, t; k, \phi, \alpha_i)}{dt}.
\end{eqnarray}

\noindent Since $\varphi$ is defined as the phase variation in
time of the perturbative wave, it is reasonable to expect constant
values of frequency, once the asymptotic state is reached.

\section{Results}

\begin{figure}
\begin{minipage}[]{0.6\columnwidth}
   \includegraphics[width=\columnwidth]{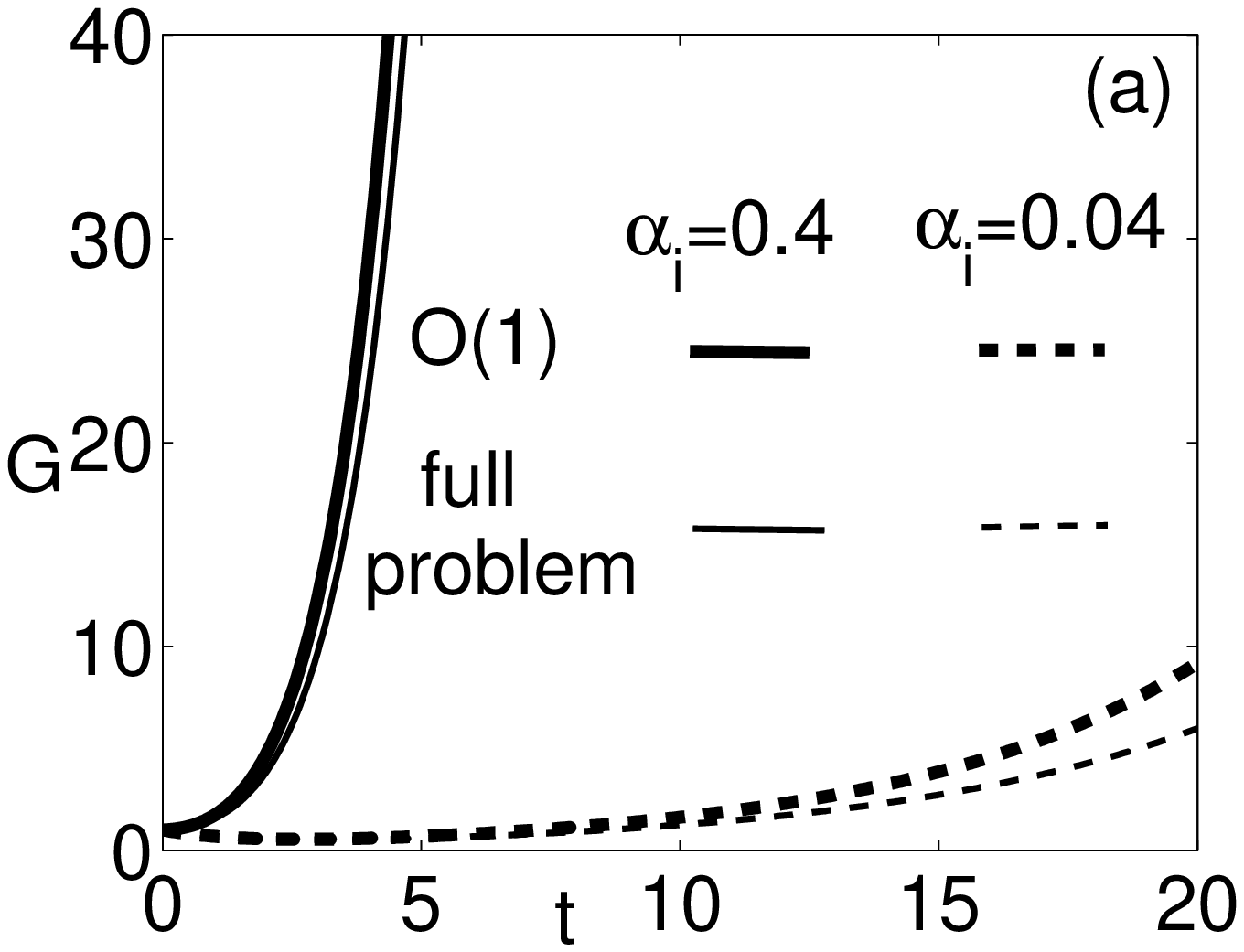}
    \label{DT_multiscale_G1}
\end{minipage}
\begin{minipage}[]{0.6\columnwidth}
   \includegraphics[width=\columnwidth]{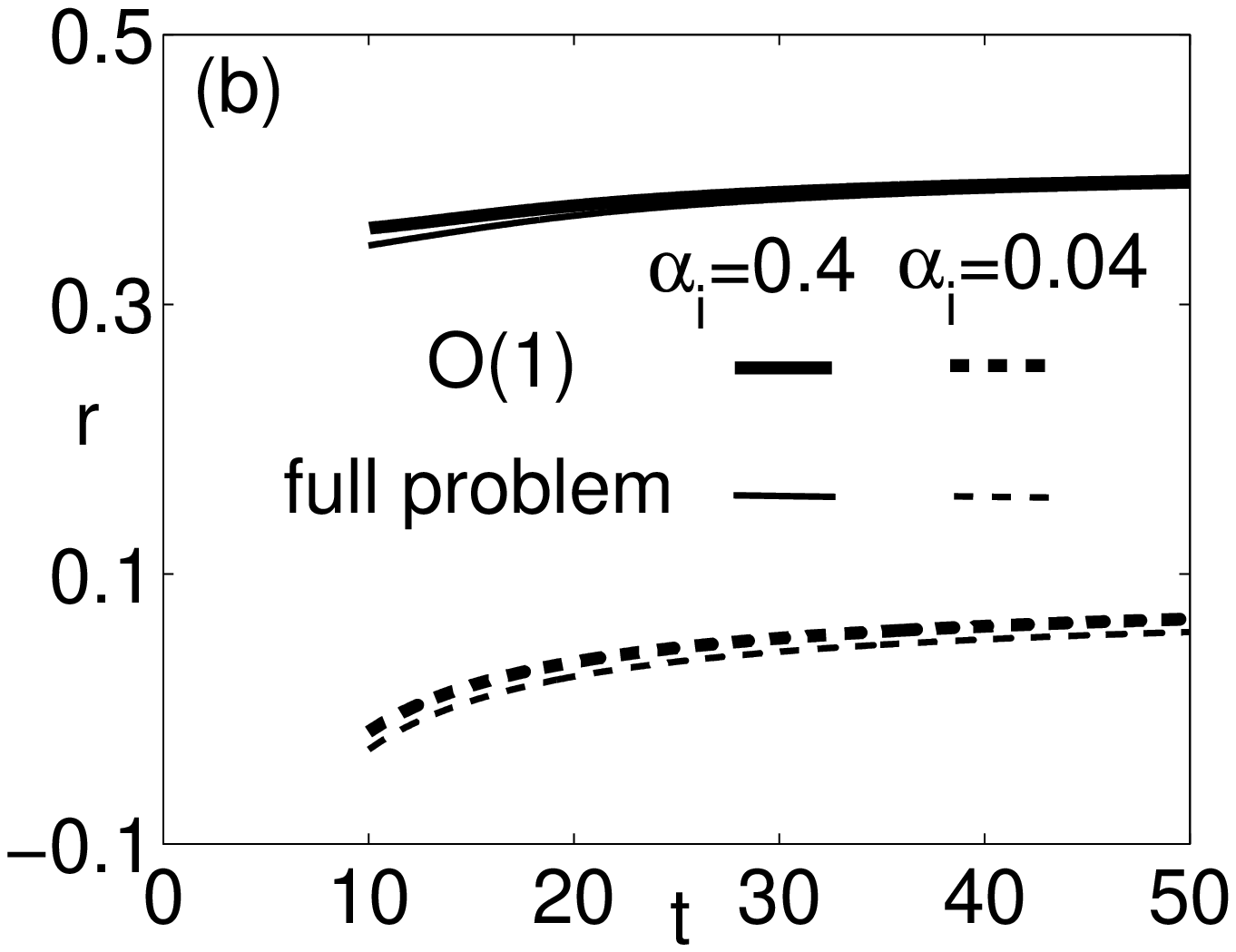}
    \label{DT_multiscale_r1}
\end{minipage}
\caption{Effects of the spatial damping rate $\alpha_i$. (a) The amplification factor $G$ and (b) the temporal growth rate $r$ as
function of time. Comparison between multiscale $O(1)$ (thick
curves) and full problem (thin curves). $Re = 50$, $k=0.03$, $\phi = \pi/4$, $x_0 = 12$, asymmetric initial
condition, $\alpha_i = 0.04, 0.4$.}\label{DT_multiscale_Gr1}
\end{figure}

Computations to evaluate the long time asymptotics are made by
integrating the equations forward in time beyond the transient
until the temporal growth rate $r$, defined in relation
(\ref{IVP2_tgr}), asymptotes to a constant value ($dr/dt <
\epsilon$) \cite{CJLJ97}, \cite{STC09}.

\begin{figure}
\begin{minipage}[]{0.6\columnwidth}
   \includegraphics[width=\columnwidth]{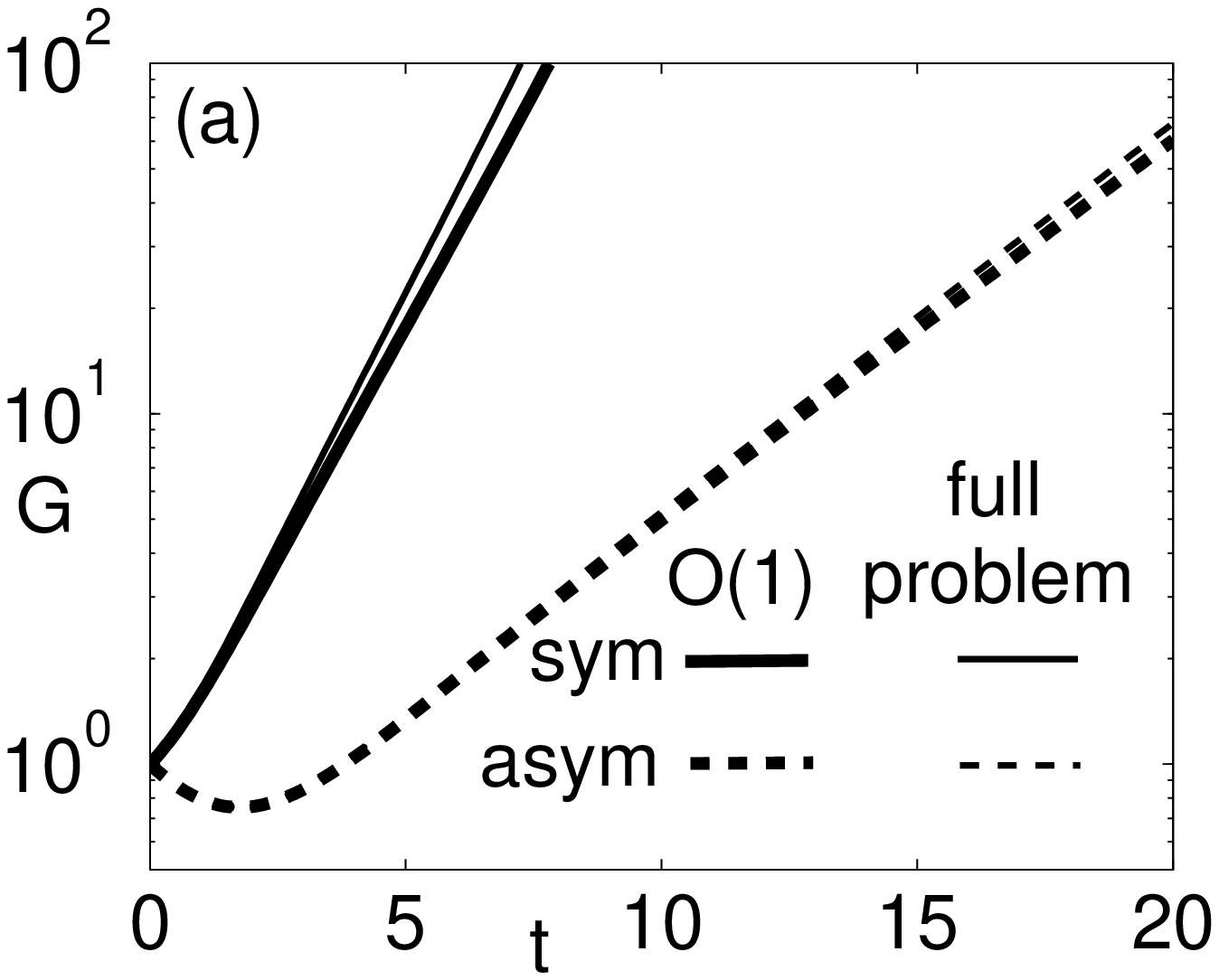}
    \label{DT_multiscale_G2}
\end{minipage}
\begin{minipage}[]{0.6\columnwidth}
   \includegraphics[width=\columnwidth]{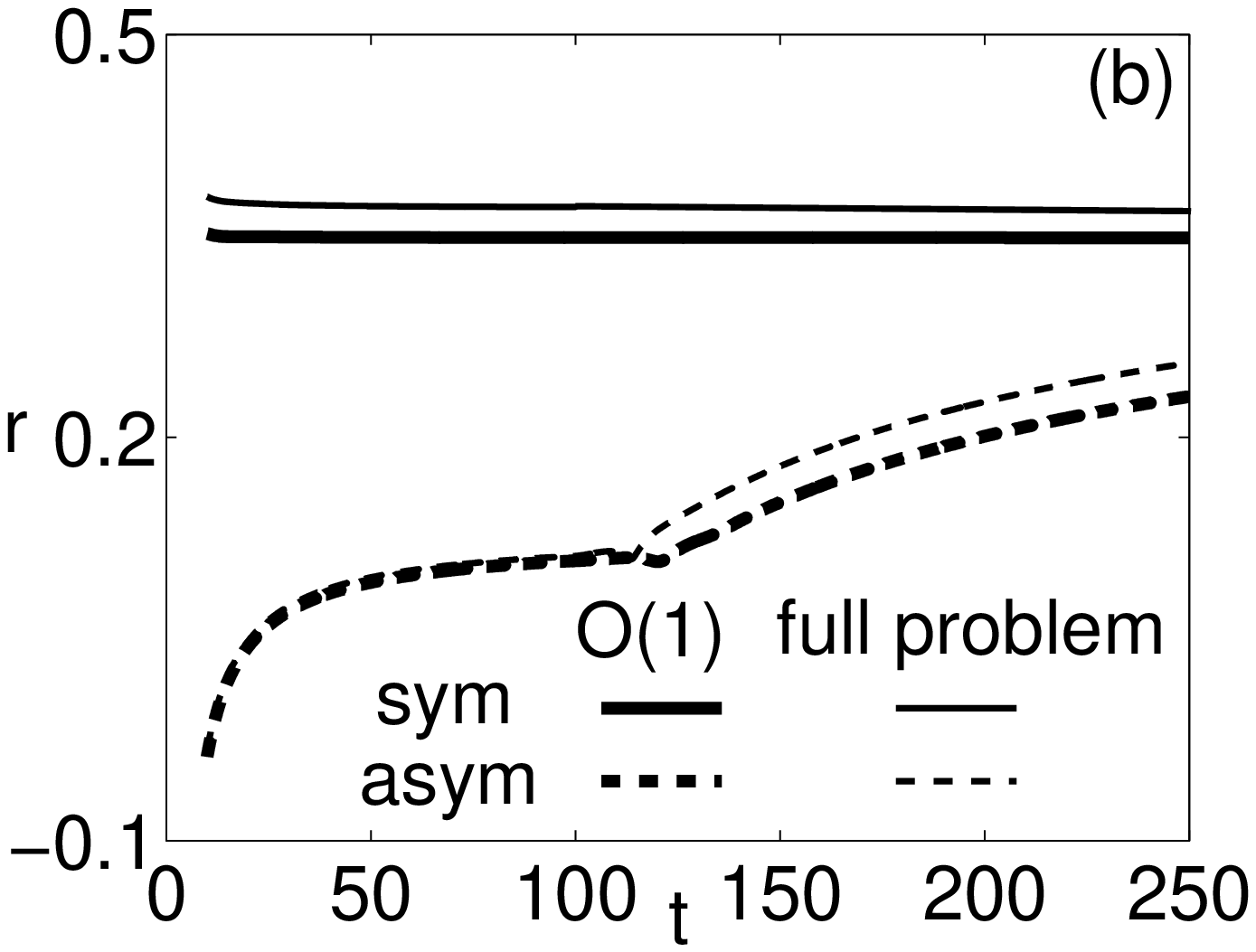}
    \label{DT_multiscale_r2}
\end{minipage}
\caption{Transversal long waves. Effect of the symmetry of the
perturbation. Comparison between multiscale $O(1)$ (thick curves)
and full problem (thin curves). (a) The amplification factor $G$ and
(b) the temporal growth rate $r$ as function of time. $Re = 100$,
$k=0.02$, $\phi = \pi/2$, $x_0 = 10$, $\alpha_i = 0.08$,
symmetric and asymmetric initial
conditions.}\label{DT_multiscale_Gr2}
\end{figure}

Fig. \ref{DT_multiscale_Gr1} presents an interesting phenomenon that
is observed for general long perturbations
(either transversal, or oblique, or longitudinal) by changing the
value of the spatial damping  $\alpha_i$. For instance, in the case
shown in this figure, which is relevant to a long oblique asymmetric
wave, the variation of the  order of magnitude of  $\alpha_i$ from
$0.04$ to $0.4$ highly enhances the amplification  in time, with a
temporal growth rate that becomes nearly three times larger.  This
means that perturbations that are spatially confined are more
amplified in time. It can also be noted that the agreement between
multiscale $O(1)$ (thick curves) and full problem (thin curves)
remains very good when changing the order  of magnitude of the
spatial damping.

\begin{figure}
\begin{minipage}[]{0.6\columnwidth}
   \includegraphics[width=\columnwidth]{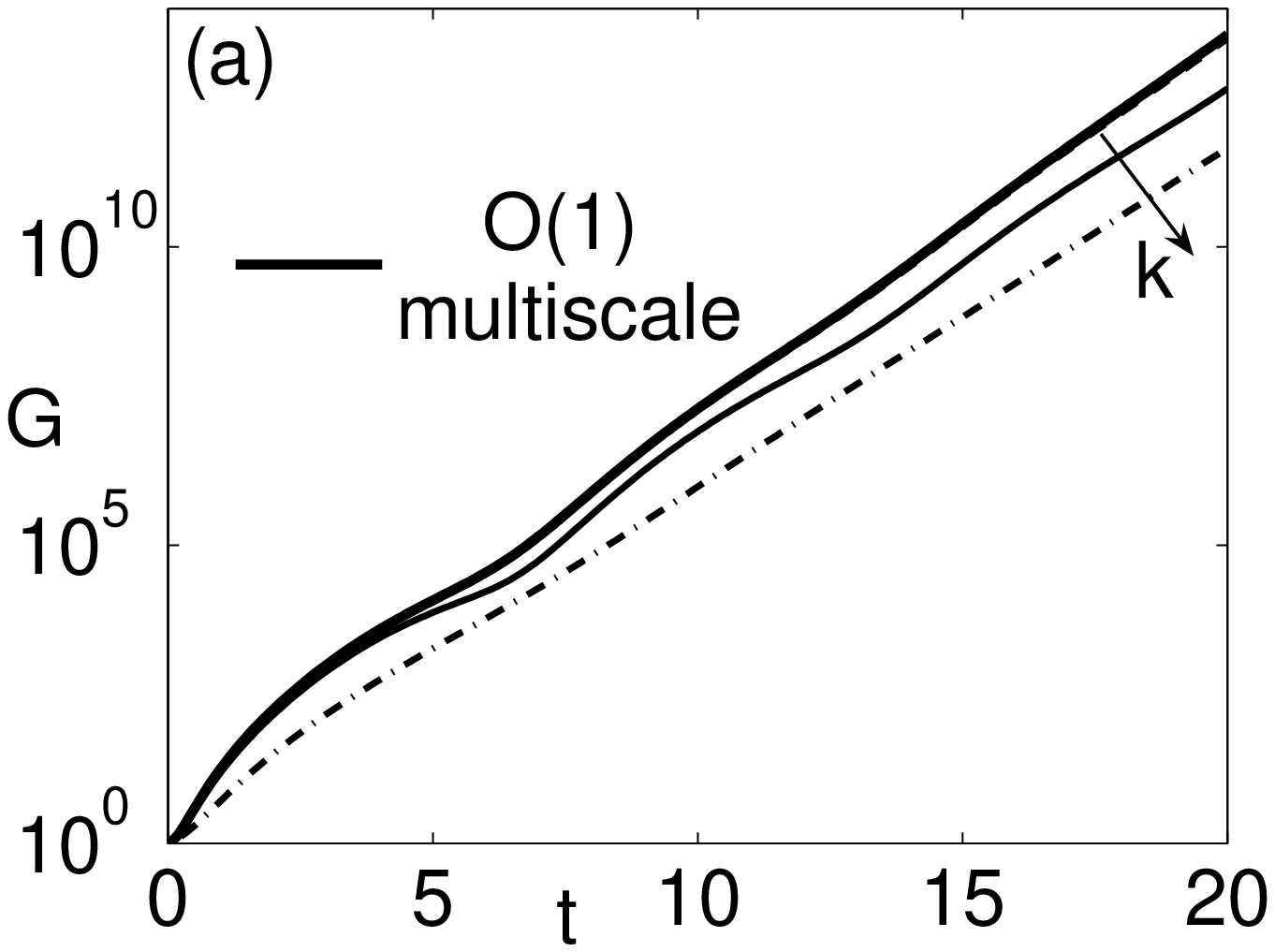}
    \label{DT_multiscale_G3}
\end{minipage}
\begin{minipage}[]{0.6\columnwidth}
   \includegraphics[width=\columnwidth]{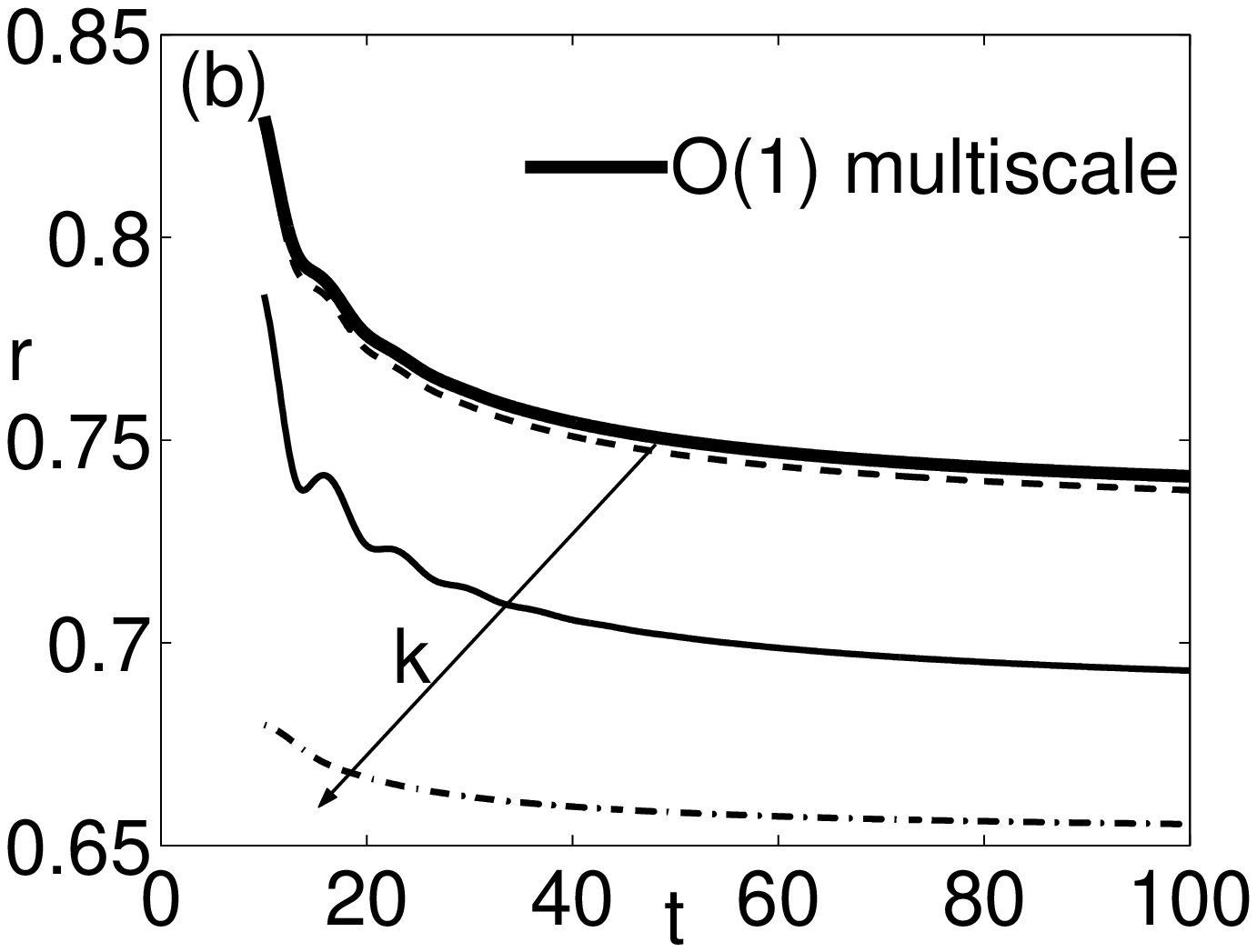}
    \label{DT_multiscale_r3}
\end{minipage}
\caption{Effects of the polar wavenumber $k$ used as small
parameter. (a) The amplification factor $G$ and (b) the temporal
growth rate $r$ as function of time. Comparison between multiscale
$O(1)$ (thick curves) and full problem (thin curves). $Re = 100$,
$\phi = 0$, $x_0 = 27$, $\alpha_i = 0.2$, symmetric initial
condition, $k = 0.1, 0.01, 0.001$.}\label{DT_multiscale_Gr3}
\end{figure}

\begin{figure}
\begin{minipage}[]{0.6\columnwidth}
   \includegraphics[width=\columnwidth]{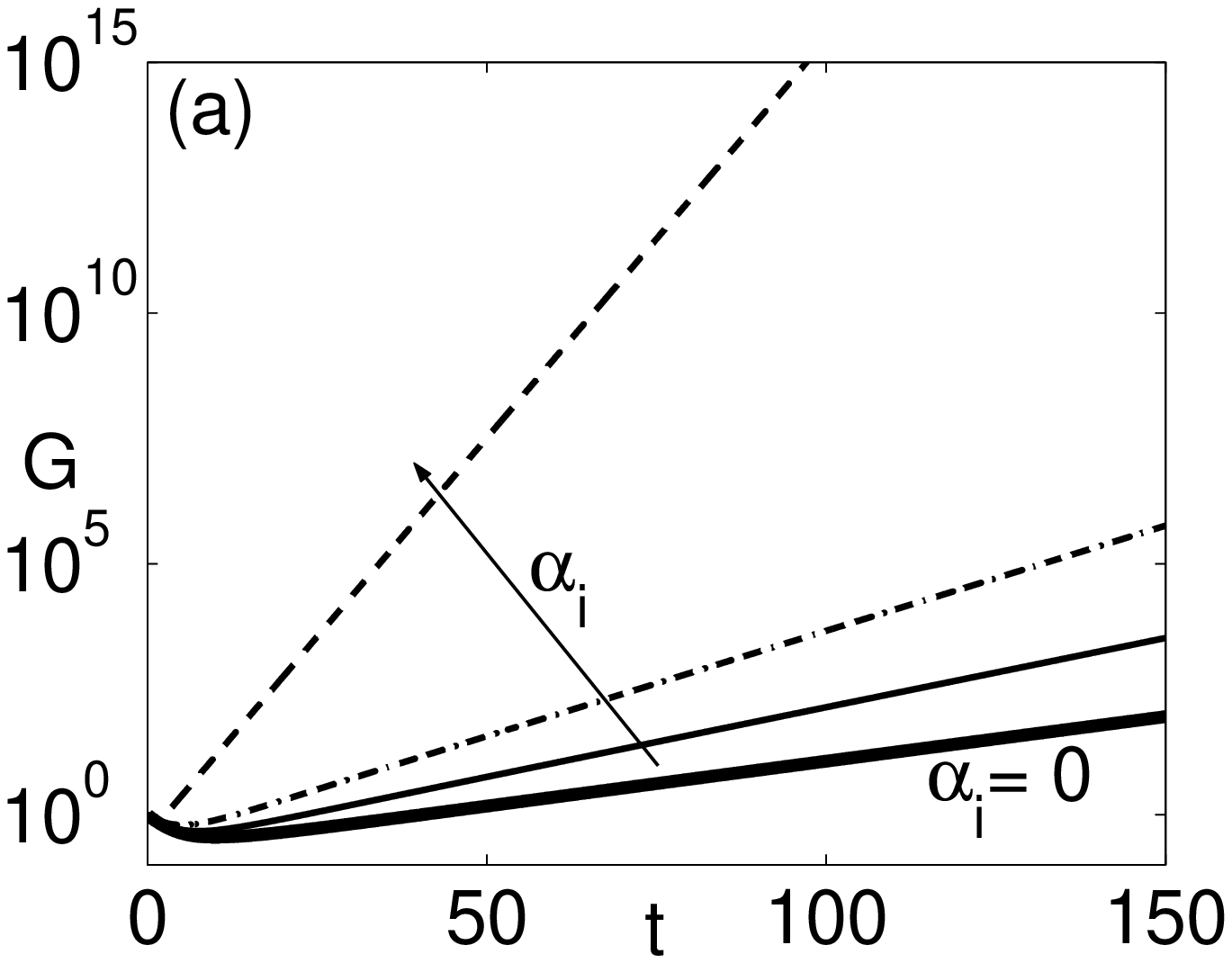}
    \label{DT_multiscale_G4}
\end{minipage}
\begin{minipage}[]{0.6\columnwidth}
   \includegraphics[width=\columnwidth]{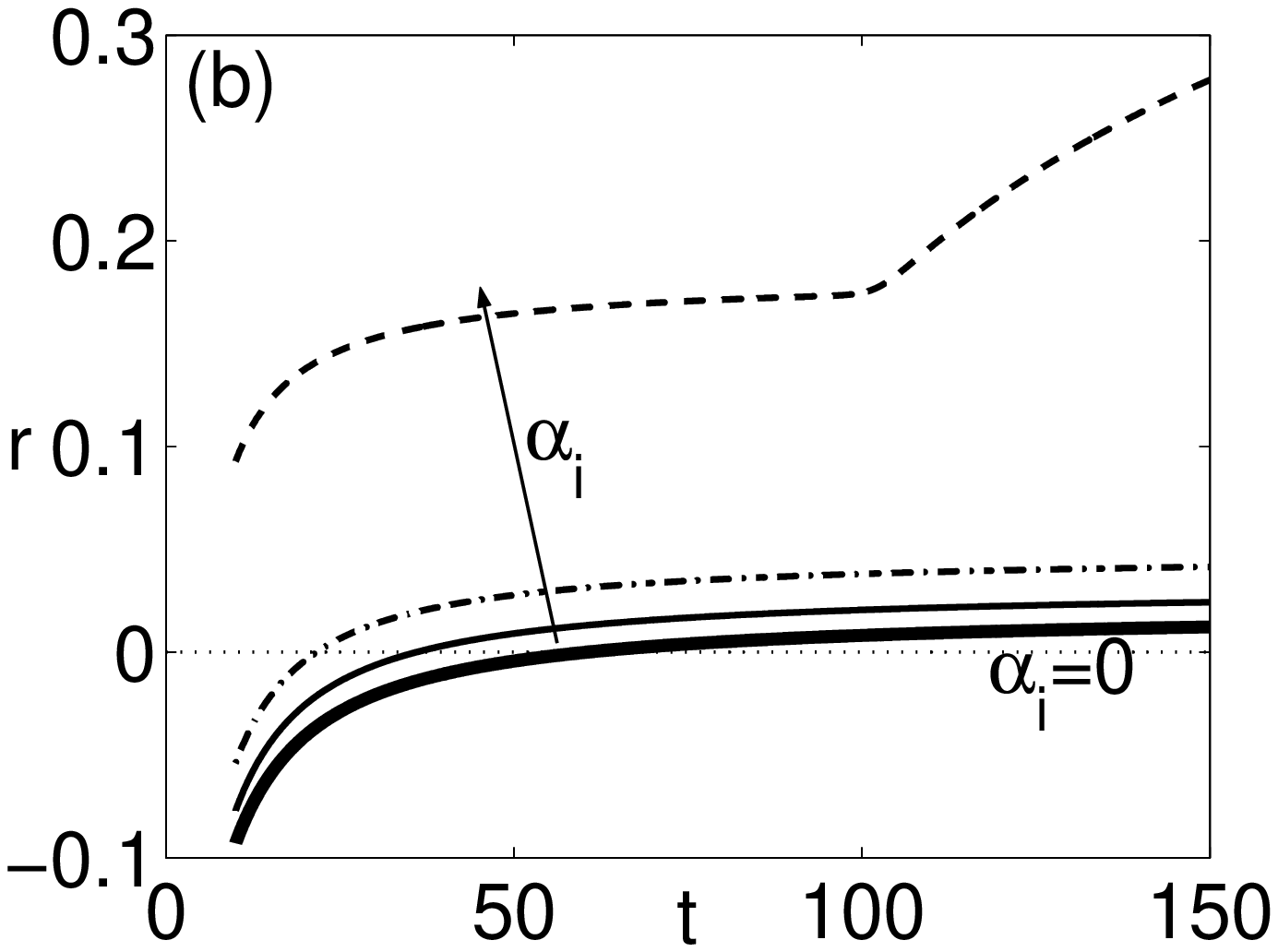}
    \label{DT_multiscale_r4}
\end{minipage}
\caption{Transversal long waves. Effect of the spatial decay,
$\alpha_i$. Comparison  between multiscale $O(1)$ (thin curves) in
the limit for $\alpha_i \rightarrow 0$, and full problem (thick
curves) with $\alpha_i = 0$. (a) The amplification factor $G$ and
(b) the temporal growth rate $r$ as function of time. $Re = 50$,
$\phi = \pi/2$, $x_0 = 10$, asymmetric initial condition, $k =
0.04$, $\alpha_i = 0.005, 0.01, 0.05$ (multiscale $O(1)$), $\alpha_i
= 0$ (full problem).}\label{DT_multiscale_Gr4}
\end{figure}

\begin{figure}
\vspace{-0.2cm}
\begin{minipage}[]{0.6\columnwidth}
   \includegraphics[width=\columnwidth]{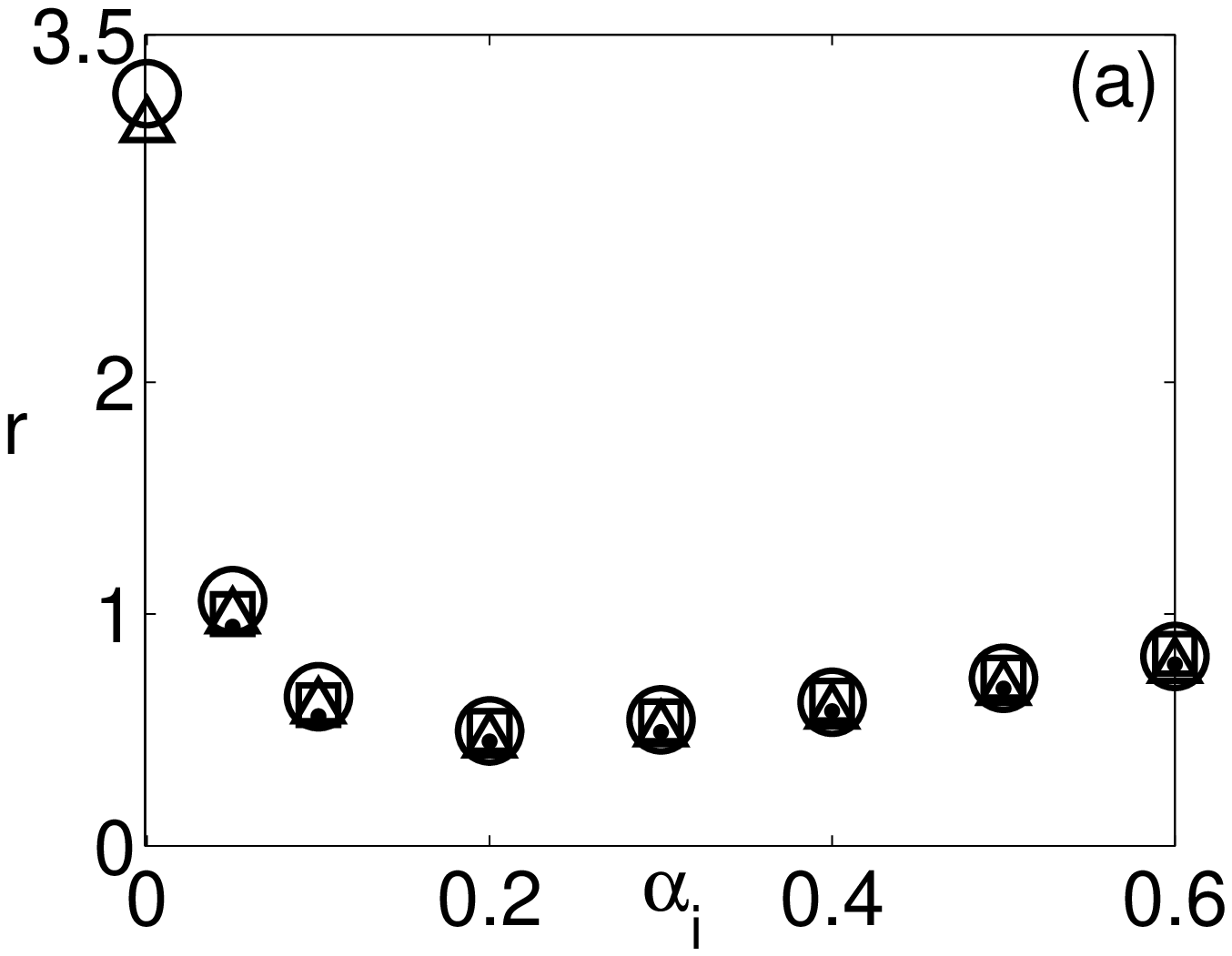}
    \label{DT_r35_full}
\end{minipage}
\vspace{-0.2cm}
\begin{minipage}[]{0.6\columnwidth}
   \includegraphics[width=\columnwidth]{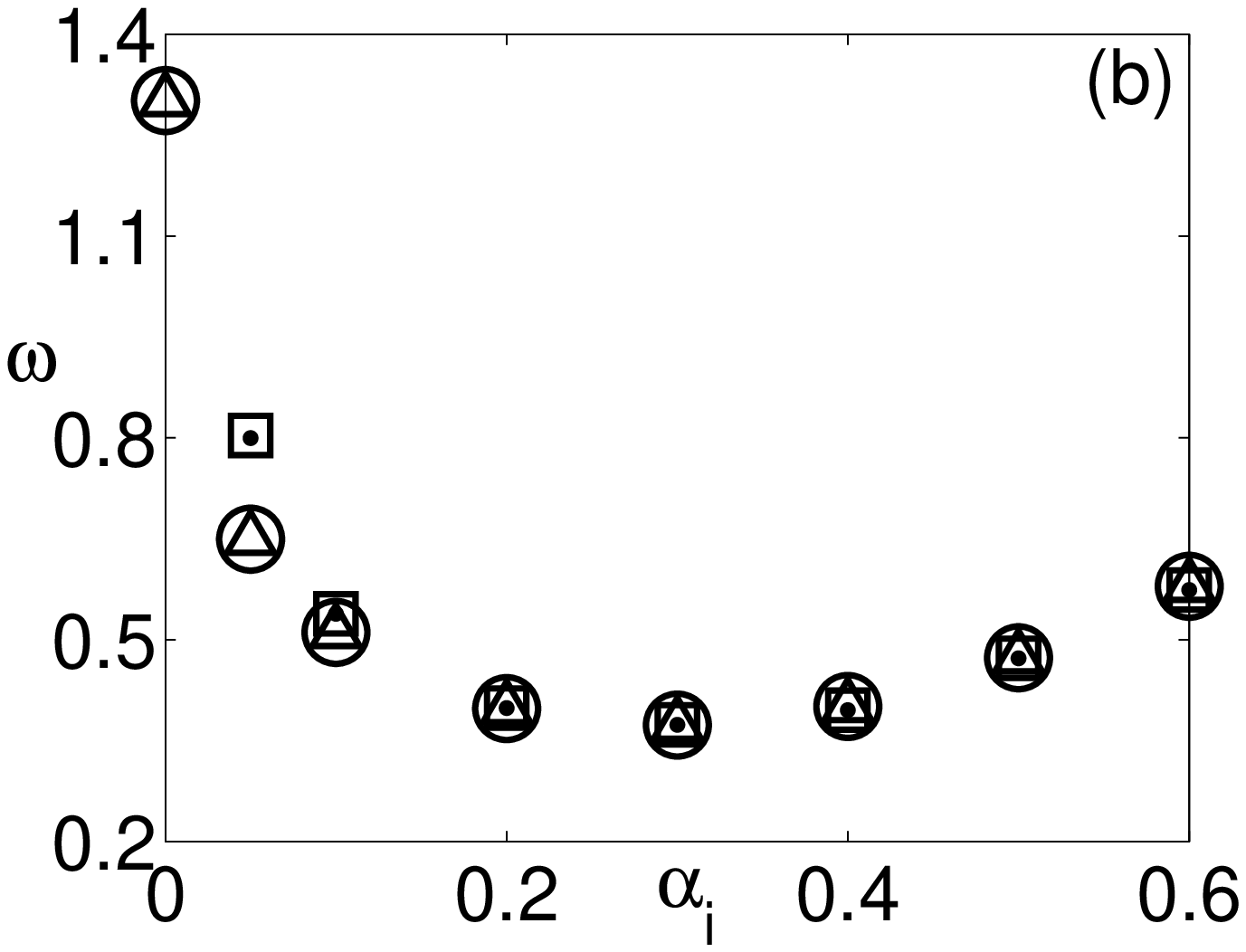}
    \label{DT_om35_full}
\end{minipage}
\caption{Temporal asymptotic values of : $\textbf{(a)}$ the temporal
growth rate and $\textbf{(b)}$ the angular frequency. Comparison
between multiscale $O(1)$ (squares: symmetric inputs, dots:
asymmetric inputs) and full linear problem (circles: symmetric
inputs, triangles: asymmetric inputs). $k = 0.01$, $Re = 100$, $\phi
= \pi/4$, $x_0=10$, $dr/dt < \epsilon$ with $\epsilon \sim
10^{-4}$.} \label{IVP_comparison_NMT_IVP_kfixed}
\end{figure}

The influence of the perturbation symmetry on the early time
behavior is shown in Fig. \ref{DT_multiscale_Gr2} (a logarithmic
scale is used on the ordinate of part (a) of the figure). It can be
noted that the symmetric initial condition leads - in the transient
behavior - to a faster temporal growth than the asymmetric one,
although both configurations are approaching the same asymptotic
unstable state. Indeed, the transient in the asymmetric case is
lasting longer than $t\sim10^2$, where in the symmetric case is
lasting $t\sim10^1$. The agreement between multiscale to $O(1)$ and
the full problem is very good for both asymmetric and
symmetric conditions. This is true both for the early transient as
well as the ultimate fate. It should be noted the discontinuous
behavior shown at  $t \sim 100$ by  the temporal growth rate of the
asymmetric transversal ($\phi=\pi/2$) wave.

The effect of differing
orders of magnitude for the polar wavenumber $k$ is highlighted in  Fig. \ref{DT_multiscale_Gr3}. Three orders are
considered, namely $k = 0.1, 0.01, 0.001$. As expected, for smaller
values of the polar wavenumber the agreement between multiscale
$O(1)$ and full problem is improving (the multiscale $O(1)$
solution practically coincides with that of the full problem for
$k = 0.001$). It is interesting to note in the interval between zero and forty base flow time scales, the presence of a temporal oscillation characterized by a period of about 10 time scales.
The transient thus presents a further  time scale beyond
its proper global one which, in this case, is lasting 100 units.

The limit for a zero spatial decay, i.e. $\alpha_i
\rightarrow 0$, was  considered in different situations, see Fig. \ref{DT_multiscale_Gr4}, transients of an orthogonal long wave perturbation, and Fig. \ref{IVP_comparison_NMT_IVP_kfixed}, time asymptotics of an oblique long wave perturbation. In Fig. \ref{DT_multiscale_Gr4}, the thick curves represent the
full problem solution with $\alpha_i = 0$, while the thin curves
are the multiscale $O(1)$ results with $\alpha_i$ values going to
zero. The right limit of the multiscale $O(1)$ solution for $\alpha_i
\rightarrow 0$ is finite, and is closely reaching the full problem
solution. As can be observed, the curves with
smaller spatial decay rates are approaching the thick curve from
above. This behavior holds in the early transient and in the
asymptotic state. 
It should be noted that, in this particular case ($Re = 50$,
$\phi = \pi/2$, $x_0 = 10$, asymmetric initial condition, $k =
0.04$),  the complete problem at $\alpha_i=0$ has a  temporal growth rate close to zero, and thus is in a near state of marginal stability. But, as previously remarked (cf. Section II), this is not a general behavior for the complete problem. However, it is true that the limiting behaviors for $\alpha_i \rightarrow  0$ of the multiscaling and of the complete problem are very close. And, as the case shown in Fig.6 confirms that, if a  difference exists, this will be located just at $\alpha_i=0$. This means that the right limit of the multiscaling for $\alpha_i = 0$ correctly approximates the complete problem, but this limit value it is not always equal to the value at $\alpha_i = 0$ (where the multiscaling yields marginal stability, i.e. $r=0, G=1, \omega=0$). It can be concluded that, in any case, the true limit of the complete problem can  be obtained by extrapolating the multiscaling results.

It is noted that, in Fig. \ref{DT_multiscale_Gr4}, for asymmetric and transversal initial conditions with a non vanishing spatial decay,  a discontinuity in the temporal growth rate can  again be observed at $t\sim 100$, see also Fig.3.

The comparison between the long waves temporal asymptotics of the full problem and its
multiscaling version is shown in Fig. \ref{IVP_comparison_NMT_IVP_kfixed}.
To consider a situation where the multiscaling applies the polar wavenumber $k$
is fixed to the value $0.01$, while the decay in space $\alpha_i$ of the longitudinal wave is in the range $[0, 0.6]$.
Multiscale to $O(1)$
results (squares and dots) are in excellent agreement, for
symmetric and asymmetric initial inputs, with full problem data
(circles and triangles). Note that the agrement improves for
increasing values of $\alpha_i$.  A minimum of the perturbation
energy (in terms of $r$) is found around $\alpha_i = 0.2-0.3$ and
a similar behavior is shown by the angular frequency $\omega$.

\section{Conclusions}

Different transient configurations have been observed by changing
the spatial damping rate, the symmetry of the perturbation and its polar wavenumber (the magnitude, in order to check the method accuracy, and the angle of obliquity).  Since the results relevant to the complete problem are in very good agreement with the results of the first order analysis, in the present work attention was
focused on the resolution of the multiscaling at order $O(1)$ only.

Two main results can be noted. First, the perturbation symmetry influences the transient. In particular, asymmetric transversal perturbations show a different kind of transient which includes an initial decay (first few time scales) and then a growth that abruptly changes its time derivative after about $100$ time scales. A sequence of such a kind of discontinuities can be envisaged up to where the growth rate of the corresponding symmetric perturbation is met. Second, the spatial decay substantially affects the transient. For example, in the case of asymmetric perturbations, it is observed that high spatial damping makes the initial temporal decay interval shorter and, at the same time, greatly increases the temporal growth rate.

Multiscale data have been compared with full problem results in the
asymptotic temporal limit. As far as small wavenumbers are
concerned, the agreement is very good for both symmetric and asymmetric initial conditions as arbitrarily
expressed in terms of elements of the trigonometrical Schauder
basis for the $L^2$ space.

Lastly, it is noted that the amplification factor of transversal perturbations never presents the trend -- a growth followed by a long damping - usually observed  in waves with wavenumber of order one or less. Asymptotically unstable configurations in time have always been observed here in the limit of long waves.

\newpage

\appendix

\section{base flow coefficients}

Here we detail the  coefficients, $\phi_i(y; x_0, Re)$  and $\chi_i(y; x_0, Re)$, of the asymptotic expansion representing the intermediate and far base flow. This approximation  is homogeneous in the $x$ and $z$ directions, and  parameterized through the downstream station $x_0$ and the Reynolds number $Re$.

\bigskip

\noindent {\bf Zero order, i=0}
\begin{eqnarray}
\phi_0 &=& C_0 \\
\chi_0 &=& 0
\end{eqnarray}

\noindent with $C_0 = 1$.

\bigskip

\noindent {\bf First order, i=1}
\begin{eqnarray}
\phi_1 &=& - A C_1 \textrm{e}^{- Re y^2/(4x_0)} \\
\chi_1 &=& 0
\end{eqnarray}

\noindent  with $C_1 = 1$.

\bigskip

\noindent {\bf Second order, i=2}
\begin{eqnarray}
\phi_2 &=& -\frac{1}{2} A^2 \textrm{e}^{- Re y^2/(4x_0)}
[C_{2}{\rm {}_1 \! F_{\!1}}(-\frac{1}{2},\frac{1}{2};\frac{Re y^2}{4x_0}) \nonumber \\
& & + \textrm{e}^{- Re y^2/(4x_0)} + \nonumber \\
&& + \frac{1}{2} \frac{y}{\sqrt x_0} \sqrt{\pi Re} \textrm{erf}(\frac{1}{2} \sqrt{\frac{Re}{x_0}} y)] \\
\chi_2 &=& - \frac{A}{2} \frac{y}{\sqrt x_0} \textrm{e}^{- Re
y^2/(4x_0)}
\end{eqnarray}

\noindent with $C_2 = - 2.75833 + 0.21237 \cdot Re
- 0.00353 \cdot Re^2 + 0.00002 \cdot Re^3$.

\bigskip

\noindent {\bf Third order,  i=3}
\begin{eqnarray}
\phi_3 &=& A^3 \textrm{e}^{- Re y^2/(4x_0)} (2 - Re
\frac{y^2}{x_0}) \times \nonumber \\
&& \times [\frac{1}{2} C_3 - Re F_3(x_0,y)] \\
\nonumber \chi_3 &=& - \frac{A^2}{2} \{C_2[-\frac{1}{2}
\frac{1}{\sqrt{x_0}} \times \nonumber \\
&& \times \int_0^y [\textrm{e}^{- Re \zeta^2/(4x_0)} {\rm {}_1 \!
F_{\!1}}(-\frac{1}{2},\frac{1}{2};\frac{Re\zeta^2}{4x_0})] d
  \zeta \nonumber \\
  & & - \frac{1}{2} \frac{y}{\sqrt{x_0}} \textrm{e}^{- Re y^2/(4x_0)}{\rm {}_1 \! F_{\!1}}(-\frac{1}{2},\frac{1}{2};\frac{Re y^2}{4x_0})] \nonumber \\
  & & - \frac{1}{2} \frac{y}{\sqrt{x_0}}
  \textrm{e}^{- Re y^2/(2x_0)}
  - \sqrt{\frac{\pi}{2Re}} \textrm{erf}(\sqrt{\frac{Re }{2 x_0}}y) \nonumber \\ & & + (\frac{1}{2}
  \sqrt{\frac{\pi}{Re}} - \frac{\sqrt{\pi Re}}{4} \frac{y^2}{x_0}) \times \nonumber \\
  && \times \textrm{e}^{- Re y^2/(4x_0)} \textrm{erf}(\frac{1}{2}
  \sqrt{\frac{Re}{x_0}}y)
\end{eqnarray}

\noindent with $C_3 = - 2.26605 + 0.15752 \cdot Re
- 0.00265 \cdot Re^2 + 0.00001 \cdot Re^3$.

Coefficient $A$ is related to the drag coefficient
$C_D$ ($A = \textstyle{\frac 1 4} (Re / \pi)^{1/2} c_D(Re)$, ${\rm
{}_1 \! F_{\!1}}$ is the confluent hypergeometric function, ${\rm
Hr}_{n-1} (\eta) = {\rm H}_{n-1} (\frac 1 2 Re^{1/2} \eta)$, where
$H_n$ are Hermite polynomials, and

\begin{eqnarray}
F_n (\eta) & = & \int \dfrac {{\rm e}^{\frac {Re} {4} \eta^2}} {{\rm Hr}_{n-1}^2 (\eta)} G_n(\eta) {\rm d} \eta ;\\
\label{FF} G_n (\eta) & = & A^{-n} \int M_n(\eta) {\rm
Hr}_{n-1}(\eta) {\rm d} \eta \; . \end{eqnarray}

\noindent where $\eta=y/\sqrt{x_0}$.

\newpage

\section{ordinary differential operators of the initial-value problem}
In this Appendix we list the ordinary differential operators of the full linear system and, up to order $O(k)$, of the  multiscale system.
The coefficients inside these operators are the quantities computed in $x_0$ that are associated to the spatial derivatives of the vorticity and velocity of the base flow ({\bf{U}}=$(U(y; x_0, Re), V(y; x_0, Re))$).  In particular, $\displaystyle{\Omega_z = ( \frac{\partial V}{\partial x}  -  \frac{\partial
U}{\partial y})  |_{x=x_0}}$ is the mean vorticity in the spanwise direction, and the coefficients $a$, $b$, $c$, $d$, $e$ are equal to: 

\noindent $\displaystyle{a= \left. \frac{\partial \Omega_z}{\partial x} \right |_{x=x_0}}$, $\displaystyle{b= \left.
\frac{\partial^2 \Omega_z}{\partial x^2} \right |_{x=x_0}}$,
$\displaystyle{c(y)= \left. \frac{\partial^2 \Omega_z}{\partial x \partial y} \right |_{x=x_0}}$, $\displaystyle{d= \left. \frac{\partial U}{\partial x} \right |_{x=x_0}}$, $\displaystyle{e= \left. \frac{\partial V}{\partial x} \right |_{x=x_0}}$.

\vspace{+0.2cm}
\noindent \textbf{Full linear problem}:

\begin{eqnarray}
\nonumber \emph{G} = &-& i (k cos(\phi) + i \alpha_i) U - V
\frac{\partial}{\partial y} +\\
&+& \frac {1} {Re}
[\frac{\partial^2}{\partial y^2} - k^2 + \alpha_i ^2 - 2
i k cos(\phi) \alpha_i], \\
\nonumber \emph{H} = &-& \frac{i (k cos(\phi) + i \alpha_i)}{k^2 + 2
i k cos(\phi) \alpha_i - \alpha_i^2} b \frac{\partial}{\partial y} -
c(y) + \\
\nonumber&-& i (k cos(\phi)
+ i \alpha_i) \frac{\partial
\Omega_z}{\partial y} + \nonumber \\
\nonumber&+& \frac{k^2 cos^2(\phi) + 2 i k cos(\phi) \alpha_i - \alpha_i^2
- k^2 sin^2(\phi)}{k^2 + 2 i k cos(\phi) \alpha_i - \alpha_i^2} \times\\
\nonumber&\times&a \frac{\partial}{\partial y} + k^2 sin^2(\phi)
\frac{\partial V}{\partial y} +
\nonumber \\
\nonumber&+& (k^2 cos^2(\phi) + 2 i k cos(\phi) \alpha_i - \alpha_i^2)
d + \nonumber \\
\nonumber&-& \frac{k^2 cos^2(\phi) + 2 i k cos(\phi) \alpha_i -
\alpha_i^2}{k^2 + 2 i k cos(\phi) \alpha_i - \alpha_i^2}
d \frac{\partial^2}{\partial y^2} +\\
\nonumber&-&\frac{k^2 sin^2(\phi)}{k^2 + 2 i k cos(\phi) \alpha_i -
\alpha_i^2} \frac{\partial V}{\partial y}
\frac{\partial^2}{\partial y^2} + \nonumber \\
\nonumber&-& i (k cos(\phi) + i \alpha_i) e
\frac{\partial}{\partial y} +\\
&+& \frac{i (k cos(\phi) + i \alpha_i)}{k^2 + 2 i k cos(\phi)
\alpha_i - \alpha_i^2} e
\frac{\partial^3}{\partial y^3}, \\
\nonumber\emph{K} = &+& \frac{k sin(\phi)}{k^2 + 2 i k cos(\phi) \alpha_i -
\alpha_i^2} b - i k sin(\phi) e +\\
\nonumber&-& 2 \frac{(k cos(\phi) + i \alpha_i) k sin(\phi)}{k^2 + 2
i k cos(\phi) \alpha_i - \alpha_i^2} a +
\nonumber \\
\nonumber&+& \frac{(k cos(\phi) + i \alpha_i) k sin(\phi)}{k^2 + 2 i
k cos(\phi) \alpha_i - \alpha_i^2} d
\frac{\partial }{\partial y} +\\
\nonumber&-& \frac{(k cos(\phi) + i \alpha_i) k
sin(\phi)}{k^2 + 2 i k cos(\phi) \alpha_i - \alpha_i^2}
\frac{\partial V}{\partial y} \frac{\partial }{\partial y} +
\nonumber \\
&-& i \frac{k sin(\phi)}{k^2 + 2 i k cos(\phi) \alpha_i -
\alpha_i^2} e \frac{\partial^2}{\partial y^2},
\end{eqnarray}

\begin{eqnarray}
\nonumber\emph{L} = &-& i (k cos(\phi) + i \alpha_i) U - V
\frac{\partial}{\partial y} +\\
\nonumber&+& \frac {1} {Re}
[\frac{\partial^2}{\partial y^2} - k^2 + \alpha_i ^2 - 2
i k cos(\phi) \alpha_i] + \nonumber \\
 &-& d + \frac{i (k cos(\phi) + i \alpha_i)}{k^2 + 2 i k
cos(\phi) \alpha_i - \alpha_i^2} e \frac{\partial}{\partial y},  \\
\nonumber\emph{M} = &-& i k sin(\phi) \frac{\partial U}{\partial y} +\\
&+&\frac{i k sin(\phi)}{k^2 + 2 i k cos(\phi) \alpha_i - \alpha_i^2}
e \frac{\partial^2}{\partial y^2}. \label{M_fullproblem}
\end{eqnarray}

\noindent \textbf{Order O(1)}:

\begin{eqnarray}
G_0 &=& \alpha_i U - V \frac{\partial}{\partial y} + \frac {1} {Re} (\frac{\partial^2}{\partial y^2} + \alpha_i^2) \label{Gh}, \\
\nonumber H_0 &=& a \frac{\partial}{\partial y} - \frac{1}{\alpha_i}
b \frac{\partial}{\partial y} +  \alpha_i \frac{\partial
\Omega_z}{\partial y} - c(y) + \nonumber \\
&-& d (\alpha_i^2 + \frac{\partial^2}{\partial y^2}) +
\frac{1}{\alpha_i} (\alpha_i^2 \frac{\partial}{\partial y} +
\frac{\partial^3}{\partial y^3}) \label{Hh}, \\
\nonumber L_0 &=& \alpha_i U - V \frac{\partial}{\partial y} + \frac {1}
{Re} (\frac{\partial^2}{\partial y^2} + \alpha_i^2) +\\
&-& d + \frac{1}{\alpha_i} e \frac{\partial}{\partial y}. \label{Lh}
\end{eqnarray}

\noindent \textbf{Order O(k)}:

\begin{eqnarray}
\nonumber G_{1} &=& - i cos(\phi) U - V \frac{\partial}{\partial Y} +\\
&+& \frac {1} {Re} [2 \frac{\partial^2}{\partial y \partial Y} - 2 i cos(\phi) \alpha_i], \label{Gh1} \\
\nonumber H_{1} &=& a \frac{\partial}{\partial Y} -
\frac{1}{\alpha_i} b \frac{\partial}{\partial Y} + \\
&-& \frac{i}{\alpha_i^2} cos(\phi) b \frac{\partial}{\partial y} - i cos(\phi) \frac{\partial \Omega_z}{\partial y} + \nonumber \\
\nonumber&-& 2 \frac{\partial^2}{\partial y \partial Y} d + 2 i
\alpha_i cos(\phi) d +\\
&+& \frac{3}{\alpha_i} \frac{\partial^3}{\partial y^2 \partial Y} e + \alpha_i e \frac{\partial}{\partial Y} + \nonumber \\
\nonumber &+& \frac{i}{\alpha_i^2} cos(\phi) e \frac{\partial^3}{\partial y^3} - i cos(\phi) e \frac{\partial}{\partial y}, \label{Hh1} \\
\nonumber K_{1} &=& \frac{2}{\alpha_i} sin(\phi) a -
\frac{i}{\alpha_i^2} sin(\phi)
b +\\
&-& i sin(\phi) e (1 - \frac{1}{\alpha_i^2}
\frac{\partial^2}{\partial y^2}), \label{Kh1}
\end{eqnarray}

\begin{eqnarray}
\nonumber L_{1} &=& - i cos(\phi) U - V \frac{\partial}{\partial Y} +\\
&+&\frac {1} {Re} [2 \frac{\partial^2}{\partial y \partial Y} - 2 i
cos(\phi) \alpha_i] + \nonumber \\
&+& \frac{1}{\alpha_i} e (\frac{\partial}{\partial Y}
+ \frac{i}{\alpha_i} cos(\phi) \frac{\partial}{\partial y}), \label{Lh1} \\
\nonumber M_{1} &=& - i sin(\phi) \frac{\partial U}{\partial y} -
\frac{i}{\alpha_i^2} sin(\phi) e \frac{\partial^2}{\partial y^2}.
\label{Mh1}
\end{eqnarray}

\newpage 

\end{document}